\documentclass[10pt, conference, compsocconf]{IEEEtran}
\usepackage{cite}
\usepackage[pdftex]{graphicx}
\usepackage{listings}
\usepackage{courier}
\usepackage{subcaption}
\usepackage{caption}
\usepackage{textcomp}

\makeatletter
\newcommand*\titleheader[1]{\gdef\@titleheader{#1}}
\AtBeginDocument{%
  \let\st@red@title\@title%
  \def\@title{%
    \bgroup\normalfont\large\centering\@titleheader\par\egroup
    \vskip1.5em\st@red@title}
}
\makeatother

\title{Towards an Area-Efficient Implementation of a High ILP EDGE Soft Processor}
\titleheader{Microsoft Research Technical Report \\ Authored January 2014; Released March 2018}

\begin{document}

\author{\IEEEauthorblockN{Jan Gray}
\IEEEauthorblockA{Gray Research LLC\\
jsgray@acm.org}
\and
\IEEEauthorblockN{Aaron Smith}
\IEEEauthorblockA{Microsoft Research\\
aaron.smith@microsoft.com}
}

\maketitle
\thispagestyle{plain}
\pagestyle{plain}
\begin{abstract}
In-order scalar RISC architectures have been the dominant paradigm in FPGA soft processor design for twenty years. Prior out-of-order superscalar implementations have not exhibited competitive area or absolute performance. This paper describes a new way to build fast and area-efficient out-of-order superscalar soft processors by utilizing an Explicit Data Graph Execution (EDGE) instruction set architecture. By carefully mapping the EDGE microarchitecture, and in particular, its dataflow instruction scheduler, we demonstrate the feasibility of an out-of-order FPGA architecture. Two scheduler design alternatives are compared.
\end{abstract}


\IEEEpeerreviewmaketitle

\section{Introduction}

Design productivity is still a challenge for reconfigurable computing. It is expensive to port workloads into gates and to endure 10\textsuperscript{2} to 10\textsuperscript{4} second bitstream rebuild design iterations. Soft processor array overlays can help mitigate these costs. The costly initial port becomes a simple cross-compile targeting the soft processors, and most design turns are quick recompiles. Application bottlenecks can then be offloaded to custom hardware exposed as new instructions, function units, autonomous accelerators, memories, or interconnects.

The advent of heterogeneous FPGAs with hard ARM cores does not diminish the complementary utility of soft cores. As FPGAs double in capacity, potential soft processors per FPGA also doubles. A mid-range FPGA can now host many hundreds of soft processors and their memory interconnection network, and such massively parallel processor and accelerator arrays (MPPAAs) can sustain hundreds of memory accesses and branches per cycle -- throughput that a few hard processors cannot match.

The microarchitectures of general purpose soft processors have changed little in two decades.
Philip Freidin's 16-bit RISC4005 (1991) was an in-order pipelined scalar RISC,
as were j32, xr16, NIOS, and MicroBlaze \cite{j32,xr16,nios,microblaze},
and as are their latest versions.
Over the years soft processors have gained caches, branch predictors, and other structures for boosting instruction level parallelism, but the basic scalar RISC microarchitecture still dominates. This reflects a good fit between this simple microarchitecture and the FPGA primitive elements required to implement it -- particularly LUTs and one write/cycle LUT RAMs. Unfortunately when such architectures take a cache miss, execution stops dead.

Design studies targeting higher instruction level parallelism (ILP)
microarchitectures typically implement VLIW~\cite{jones:2005,tilt:2013}
or vector~\cite{vespa:2008,yu:2008} architectures instead of
out-of-order (OoO)~\cite{mit:mips,cfc09,fpgasim:2012} soft processor cores. The
problem with superscalar OoO microarchitectures is the complexity
of the machinery needed to rename registers, schedule instructions
in dataflow order, clean up after mispeculation, and retire
results in-order for precise exceptions. This in turn requires
expensive circuits such as deep many-ported register files,
many-ported CAMs for dataflow instruction scheduling wakeup, and
many wide bus multiplexers and bypass networks, all of which are
area intensive in FPGAs. For example, multi-read, multi-write RAMs
require a mix of replication, multi-cycle operation, clock doubling,
bank interleaving, live-value-tables, and other expensive techniques.

\begin{figure}[t!]
\centerline{
  \includegraphics[width=3.1in]{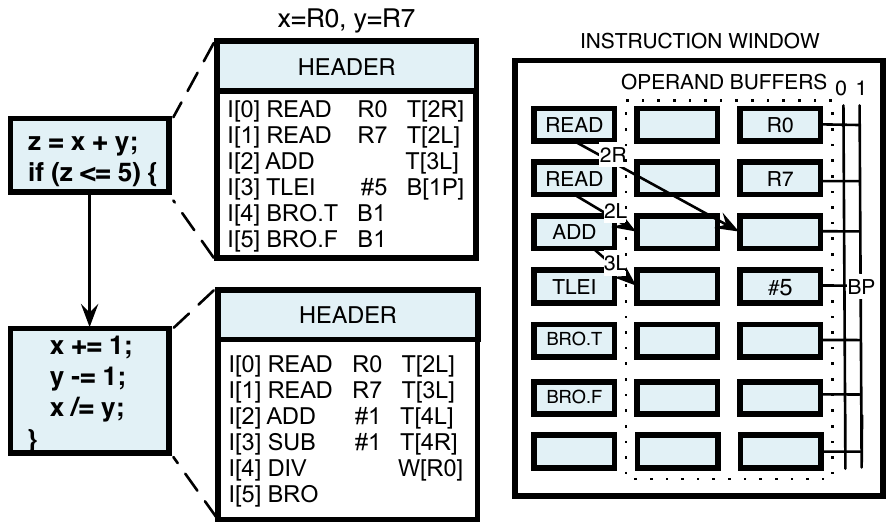}
}
\caption{Psuedo code and corresponding instruction block.}
\label{fig:edge}
\end{figure}

The present work is a new approach to build high ILP
OoO superscalar soft processors without most of the complexity and
overhead. Our insight is to implement an Explicit Data Graph
Execution (EDGE)~\cite{burger:computer04,asplostrips} instruction set architecture
designed for area and energy efficient high ILP execution. Together
the EDGE architecture and its compiler finesse away much of the register
renaming, CAMs, and complexity, enabling an out-of-order processor
for only a few hundred LUTs more than an in-order scalar RISC.

This paper explores how an EDGE ISA and 
FPGA optimized EDGE microarchitecture compare to in-order RISCs common on FPGAs today.
The key challenge, and the main contribution of the paper, is how
to build a small, fast dataflow instruction scheduler in an FPGA.
We develop and contrast two alternative FPGA implementations
on our way to developing a minimal-area EDGE soft processor.

\section{EDGE Overview}

EDGE architectures~\cite{kim07,burger:computer04,robatmili:2013,sibi:2013} execute instructions organized within instruction blocks that are fetched, executed, and committed atomically. Instructions inside blocks execute in dataflow order, which removes the need for expensive register renaming and provides power efficient out-of-order execution. The compiler explicitly encodes the data dependencies through the instruction set architecture, freeing  the microarchitecture from rediscovering these dependencies at runtime. Using predication, all intra-block branches are converted to dataflow instructions, and all dependencies other than memory are direct data dependencies. This {\it target form} encoding allows instructions within a block  to communicate their operands directly via operand buffers, reducing the number of accesses to a power hungry multi-ported physical register file. Between blocks, instructions communicate using memory and registers. By utilizing a hybrid dataflow execution model, EDGE architectures still support imperative programming languages and sequential memory semantics, but reap the benefits of out-of-order execution with near in-order power efficiency and complexity.

Figure~\ref{fig:edge} shows an example of two EDGE instruction blocks and how instructions explicitly encode their targets. In this example each block corresponds to a basic block. The first two READ instructions target the left (T[2L]) and right (T[2R]) operands of the ADD instruction. A READ is the only instruction that reads from the global register file (however any instruction may target, i.e. write to, the global register file). When the ADD receives the result of both register reads it will become ready and execute. 

\begin{figure}[t!]
\centering
\includegraphics[width=3.2in]{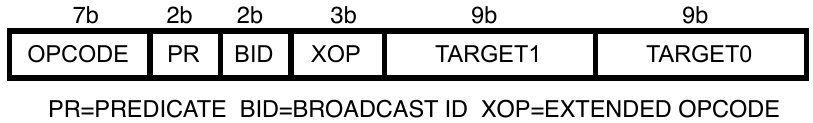}
\caption{General instruction format}
\label{fig:instruction-format}
\end{figure}

Figure~\ref{fig:instruction-format} shows the general instruction format. Each EDGE instruction is 32 bits and supports encoding up to two target instructions. For instructions with more consumers than target fields, the compiler can build a fanout tree using move instructions or it can can assign high fanout instructions to broadcasts~\cite{robatmili:2013}. Broadcasts support sending an operand over a lightweight  network  to any number of consumer instructions in a block. In Figure~\ref{fig:edge}, when the TLEI (test-less-than-equal-immediate) instruction receives its single input operand from the ADD it will become ready and execute. The test then produces a predicate operand that is broadcast on channel one (B[1P]) to all instructions listening on the broadcast channel, which in this example are the two predicated branch instructions (BRO.T and BRO.F). The branch that receives a matching predicate will fire.

A spectrum of EDGE implementations are possible with various area and performance tradeoffs.
Prior EDGE research studied very wide issue implementations~\cite{burger:computer04,asplostrips},
as well as fusion of multiple cores~\cite{kim07,robatmili:2013,sibi:2013}  to boost performance on scalar workloads.
In this work we focus on MPPAA scenarios utilizing compact EDGE soft processors with competitive performance/area.
Therefore data and pointers are 32 bits; blocks can be up to 32 instructions
long; and the microarchitecture decodes 1-2 instructions per clock and issues one.
We further restrict the load-store queue (LSQ) in this study to a simple, non-speculative design and omit
branch or memory dependence prediction.

\section{EDGE in an FPGA}
\subsection{Microarchitecture}

\begin{figure}[ht]
\centerline{
  \includegraphics[width=2.8in, trim=0.25in 0in 0.4in 0in]{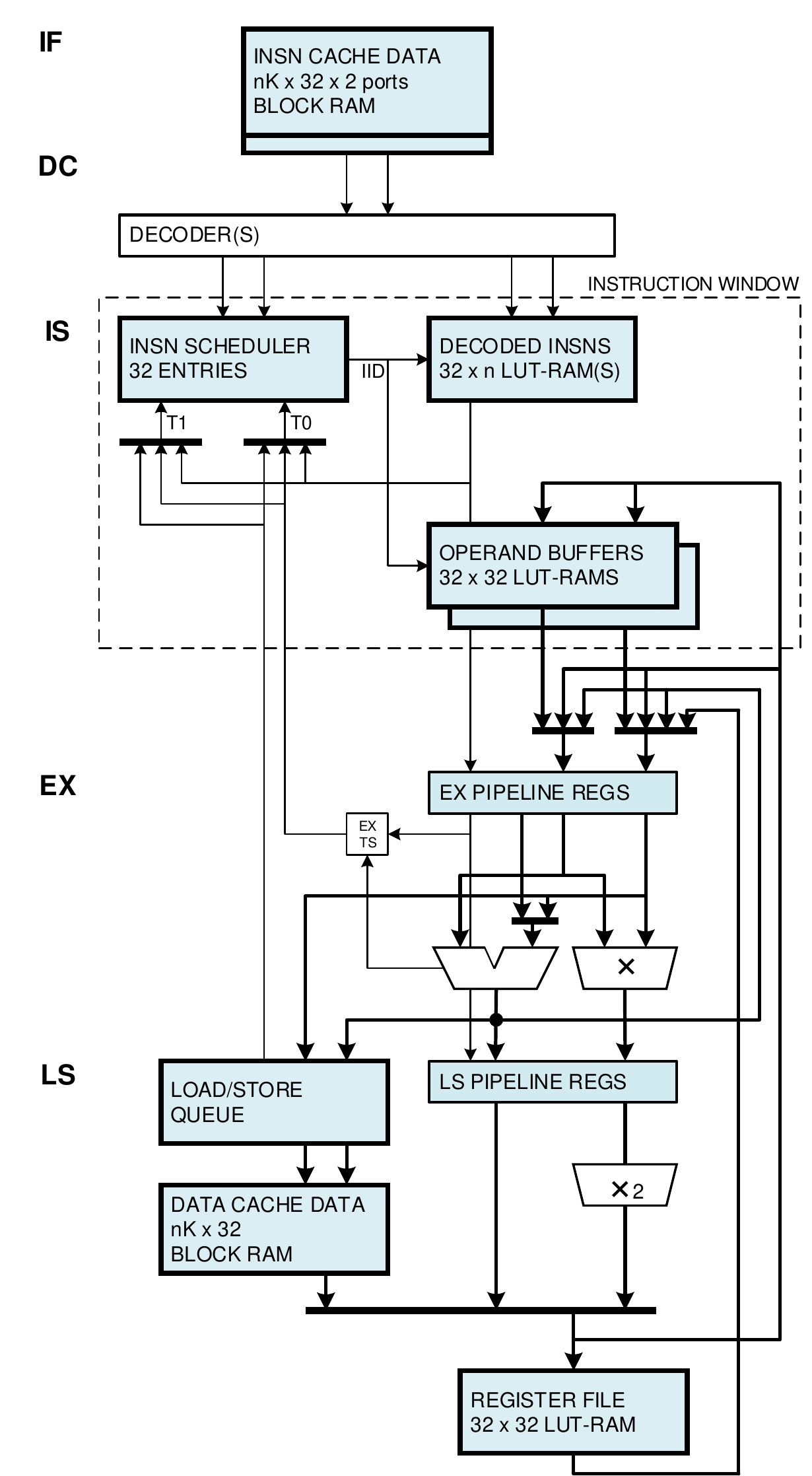}
}
\caption{Two decode, single issue EDGE microarchitecture.}
\label{fig:e2xs}
\end{figure}

Figure~\ref{fig:e2xs} is an example microarchitecure for a compact EDGE processor. It has much in common with a conventional in-order scalar RISC: instruction and data caches and a five stage pipeline including instruction fetch (IF), decode (DC), operand fetch, execute (EX), and memory/data cache access (LS). Unlike an in-order processor, instruction operands are read from operand buffers, not the register file; and the instruction to execute next, in dataflow order, is determined by the IS (issue) pipeline stage. This employs an instruction window comprising a dataflow instruction scheduler, a decoded instructions buffer, and operand buffers. It uses a simple load-store queue to issue memory instructions in program order.

The front end (IF, DC) runs decoupled from the back end (IS, EX, LS). It fetches and decodes two instructions per clock into the instruction window.
The instruction window's dataflow scheduler keeps the {\it ready state} of each decoded instruction's inputs i.e. its predication and operands. When all of its inputs (if any) are ready, the instruction wakes up and is ready to issue. The lowest numbered ready instruction {\it IID} is selected each cycle and its decoded instruction and input operands are read. Besides the  data mux and function unit control signals, this instruction encodes up to two {\it ready events}. The scheduler accepts these and/or events from other sources (muxed into {\it T0} and {\it T1}) and updates the ready state of other instructions in the window. Thus dataflow execution proceeds, starting with the block's ready 0-input instructions, then instructions that these target, and so forth.

\subsection{EDGE dataflow instruction scheduling requirements}
The instruction window and scheduler are the linchpin of the core. Their area, clock period, capabilities, and limitations largely determine the realized performance of an EDGE core and the throughput of an EDGE multiprocessor.

The instruction scheduler has diverse functionality and requirements.
It is highly concurrent.
Each cycle the decoder(s) write instructions' {\it decoded ready state} and decoded instructions into the window.
Each cycle the scheduler selects the next instruction to issue, and in response the back end sends {\it ready events} -- either {\it target ready events} targeting a specific instruction's input slot (predicate, operand \#0, operand \#1) or {\it broadcast ready events} targeting all instructions waiting on a broadcast ID.
These set per-instruction {\it active ready state} bits which together with the decoded ready state may signal that the instruction is ready to issue.
Note the scheduler sometimes accepts events for target instructions which have not yet been decoded and
must also inhibit reissue of issued ready instructions.

EDGE instructions may be non-predicated, or predicated true or false. A predicated instruction does not become ready until it is targeted by another instruction's predicate result, and that result matches the predicate condition. If the predicate doesn't match, the instruction never issues.

On branch to a new block all instruction window ready state is flash cleared ({\it block reset}).
However when a block branches back to itself ({\it block refresh})
only active ready state is cleared; the decoded ready state is preserved so that
it is not necessary to re-fetch and decode the block's instructions.
Refresh is key to saving time and energy in loops.

Since some  software critical paths consist of a single chain
of dependent instructions, i.e. A targets B targets C, it is important
that the dataflow scheduler add no pipeline bubbles for successive
back-to-back instruction wakeup.
Therefore the IS-stage ready-issue-target-ready pipeline recurrence should complete in one cycle
-- assuming this does not severely impact clock frequency.

Instructions such as ADD have a latency of one cycle.
With EX-stage result forwarding the scheduler can wake their targets' instructions in the IS-stage,
even before the instruction completes.
Other instruction results may await ALU comparisons, take multiple cycles,
or have unknown latency.
These must wait until later to wake their targets.

Finally, the scheduler design should be scalable across a spectrum of anticipated
EDGE implementations -- each cycle accepting at least 1-4 decoded instructions
and 2-4 target ready events, and issuing 1-2 instructions per cycle.

We consider two alternative dataflow instruction scheduler designs: a brute-force {\it parallel} scheduler, where instructions' ready state is explicily represented in FPGA D-flip-flops (FFs), in which the ready status of every instruction is reevaluated each cycle; and a more compact {\it incremental} scheduler which keeps ready state in LUT RAM and which updates ready status of only 2-4 target instructions per cycle.

\subsection{A parallel instruction scheduler}
\begin{figure}[ht]
\centerline{
  \includegraphics[width=3in, clip=true, trim=0in 0.25in 0in 0.4in]{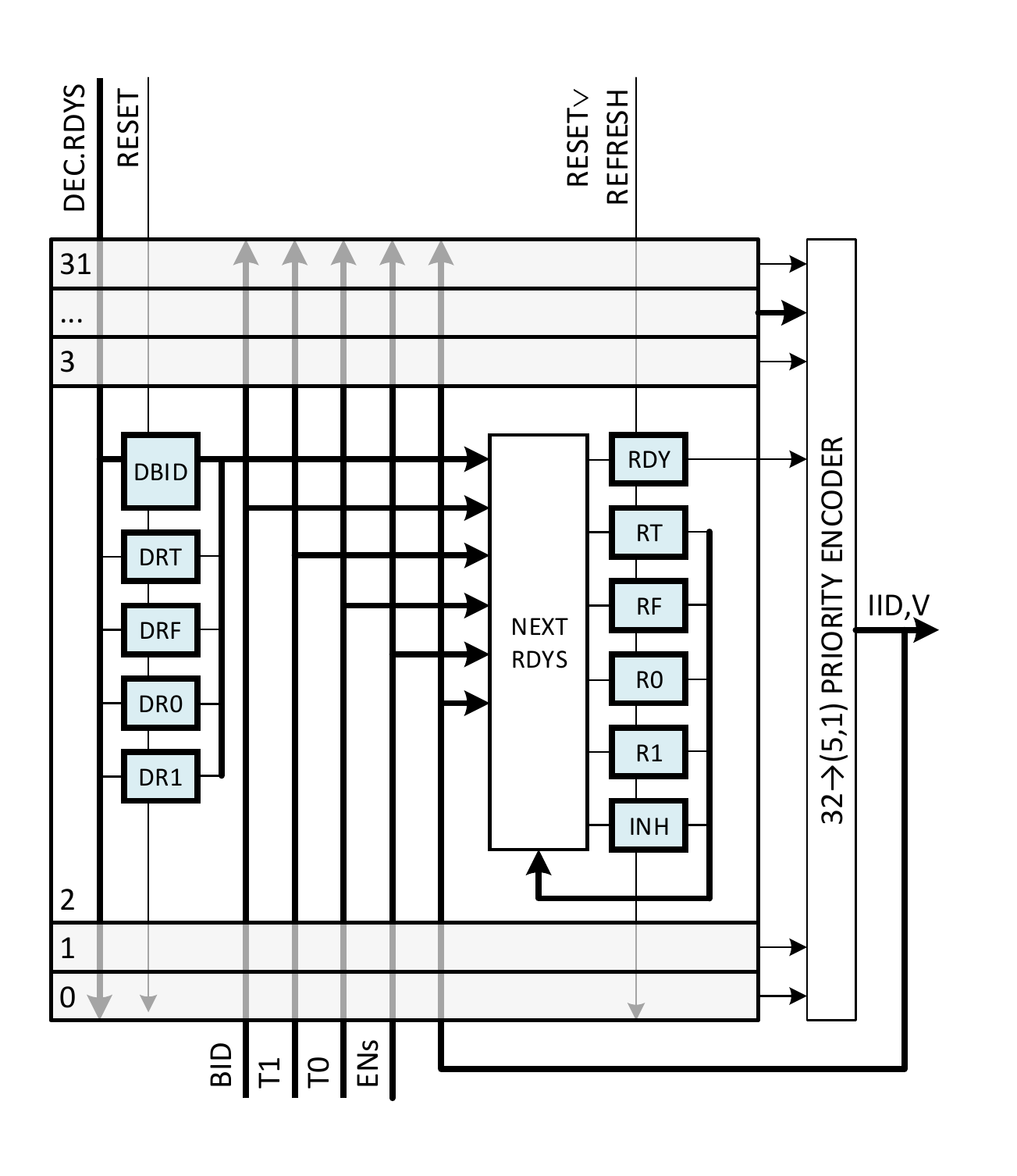}
}
\caption{Block diagram of a parallel dataflow scheduler, with entry \#2 shown in more detail.}
\label{fig:parallel}
\end{figure}

Figure~\ref{fig:parallel} depicts a parallel instruction scheduler for the
instruction window of Figure~\ref{fig:e2xs}.
Note the active ready state is set by target ready events {\it T0, T1} and broadcast ID {\it BID} (if any),
qualified by various input type enables {\it ENs}.
For a 32-entry window there are 32 instances of a one-instruction ready circuit.
In any cycle one or more of the 32 {\it RDY} signals may be asserted.
A 32-bit priority encoder reduces this to the 5-bit {\it IID} of the next instruction to issue.

For each entry there are six bits of decoded ready state,
i.e. they are initialized by the instruction decoder:
\begin{itemize}
  \item {\it DBID}: 2-bit binary broadcast ID, or 00 if none
  \item {\it DRT, DRF}: decoder: predicate true (false) is ready
  \item {\it DR0, DR1}: decoder: operand \#0 (operand \#1) is ready
\end{itemize}
Together these bits encode whether the instruction has been decoded, awaits a predicate and/or some operand(s),
perhaps via a broadcast channel, or is immediately ready to issue.
These bits are cleared on block {\it reset} only.

There are also six bits of active ready state:
\begin{itemize}
  \item {\it RT, RF}: predicate true (false) is ready
  \item {\it R0, R1}: operand \#0 (operand \#1) is ready
  \item {\it INH}: inhibit instruction -- it has already issued
  \item {\it RDY}: instruction is ready to issue
\end{itemize}
An instruction is ready iff {\it (RT \& RF \& R0 \& R1 \& \texttildelow INH)}.
Any of {\it RT, RF, R0, R1} may be set when:
\begin{itemize}
  \item{its corresponding {\it DRX} is set by the decoder, or}
  \item{an executing instruction targets that input, explicitly, or via a broadcast event {\it (broadcast ID, input)}.}
\end{itemize}
Active ready state bits are cleared on block {\it reset} or {\it refresh}.

\begin{table}[bht]
\setlength{\tabcolsep}{2pt}
\centering
\begin{tabular}{|l||c|c|c|c|c||c|c|c|c|c|c|}
\hline
&\multicolumn{5}{|c||}{Decoded ready state}&\multicolumn{6}{|c|}{Active ready state}\\
\hline
Instruction&DBID&DRT&DRF&DR0&DR1&RT&RF&R0&R1&INH&RDY\\
\hline
READ&00&1&1&1&1&1&1&1&1&{\bf 1}&0 \\
READ&00&1&1&1&1&1&1&1&1&0&{\bf 1} \\
ADD&00&1&1&0&0&1&1&{\bf 1}&0&0&0 \\
TLEI&00&1&1&0&1&1&1&0&1&0&0 \\
BRO.T B1&{\bf 01}&{\bf 0}&1&1&1&0&1&1&1&0&0 \\
BRO.F B1&{\bf 01}&1&{\bf 0}&1&1&1&0&1&1&0&0 \\
{\it undecoded}&00&{\bf 0}&{\bf 0}&x&x&0&0&x&x&x&0 \\
\hline
\end{tabular}
\caption{Example Instruction Scheduler Ready State}
\label{tab:ready-state}
\end{table}
Table~\ref{tab:ready-state} depicts a block's instruction scheduler state after decoding
six instructions and issuing the first.
The first four non-predicated instructions have {\it DRT} and {\it DRF} set reflecting that
they do not await any particular predicate results.
The two READ instructions, unpredicated and with zero input operands, are immediately ready to issue.
The first has issued -- and so is now inhibited from reissue -- targeting operand 0 of the ADD, whose {\it R0} is now set.
The second READ will issue in the next IS pipeline cycle.
The TLEI ({\it test-less-than-or-equal-immediate}) instruction broadcasts its predicate outcome on channel 1;
the two branch instructions, predicated true (resp. false), await this predicate result.
The seventh entry has not been decoded: {\it (DRT\textbar DRF)=0}.

To reduce the critical path of dataflow scheduling, the front end writes predecoded
EDGE instructions into the decoded instructions buffer.
As instruction {\it IID} issues, its decoded instruction is read by the back end.
Amongst other things it contains two target operand ready event fields, {\it \_T0} and {\it \_T1},
which designate the 0-2 {\it (IID, input)} explicit targets of the instruction, as well as 
a 4-bit vector of input enables: {\it ENs=\{RT\_EN, RF\_EN, R0\_EN, R1\_EN\}}.
Referring back to Figure~\ref{fig:e2xs}, these signals are muxed with ready events
from other pipeline stages into {\it T0} and {\it T1} input by the scheduler.

\subsection{FPGA implementation of the parallel scheduler}
Careful attention to FPGA circuit design is required to
minimize the area and clock period of the scheduler.
The 32-instruction window requires 32*(6+6)=384 FFs for the ready state,
and 32*{\it many} LUTs to decode ready events and update each entry's ready state.

A modern FPGA packs a set of LUTs (lookup tables) and D-flip-flops (FFs) together into a logic cluster.
For example, Xilinx 7 series devices group four 6-LUTs and eight FFs into each ``slice'' cluster.
Each LUT has two outputs and may be used as one 6-LUT, or two 5-LUTs with five common inputs.
Each output may be registered in a FF.
The flip-flops have optional CE (clock enable) and SR (set/reset) inputs but
these signals are common to all eight FFs in the cluster.
This basic cluster architecture is similar in Altera FPGAs.

From this follows two design considerations.

{\it Fracturable 6-LUT decoders:}
For target instruction index decoding, so long as indices are $\le$5 bits, two decoders may fit into a single 6-LUT.

{\it Slice FF packing and cluster control set restrictions:}
To minimize area and wire delays, the design packs the ready state FFs densely, 4-8 FFs per cluster.
Every 6-bit decoded ready state entry is written together (common RST and CE)
and can pack into one or two slices.

More care is required for the active ready state FFs.
Each of these 32*6=192 FFs may be individually set, but by packing four FFs per slice,
when one FF is clock enabled, all are clock enabled.
Whenever a FF is set by a ready event the other FFs in its slice should not change.
This requires implementing CE functionality in each FF's input LUT, feeding back its output into its input:
{\it FF\_NXT~=~FF~\textbar~(EN~\&~input)}.

Listing~\ref{nextreadys} is Verilog that generates the ``next readys`` for an {\it N}-entry parallel scheduler.
Although there are four ready event input types (predicate true, false, operand \#0, operand \#1),
by ensuring that {\it predicate} target events never occur in the same cycle as {\it operand} target events, a single target index bit
suffices to distinguish false/operand \#0 targets from true/operand \#1 targets.
(Further decoding is provided by specific {\it \{RT/RF/R0/R1\}\_ENs} enables.)
Therefore for an instruction window with {\it N=32} entries, T0 and T1 are six bits {\it \{input\#:1; IID:5\}}.
The target decoders {\it T00, T01, T10, T11} (target-0-input-0, etc.) are each one 6-LUT, as is the
broadcast select decoder {\it B}.
The next active ready state logic folds together the target decoder outputs with current active and decoded ready state.
This requires another seven LUTs (two for INH\_NXT), for a total of 32*12 = 384 LUTs.

This may be improved by splitting the 32-entry scheduler into {\it two 16-entry banks}
of even and odd instructions.
Within a bank a 4-bit {\it bank-IID} suffices.
Then {\it T0, T1} narrow to five bits so {\it T00, T01, T10, T11} fit in two 5,5-LUTs, and {\it INH\_NXT} in one 6-LUT,
or 2*16*(3+6)=288 LUTs in all.

\begin{lstlisting}[label=nextreadys,caption=Parallel scheduler ``next readys'' logic]
generate for (i = 0; i < N; i = i + 1) begin: R
  always @* begin
    // target decoders
    T00[i] = T0 == i;
    T01[i] = T0 == (i|N);
    T10[i] = T1 == i;
    T11[i] = T1 == (i|N);
    B[i]   = BID == DBID[i];

    // next active ready state logic
    RT_NXT[i]  = RT[i] | DRT[i]
               | (RT_EN & (T01[i]|T11[i]|B[i]));
    RF_NXT[i]  = RF[i] | DRF[i]
               | (RF_EN & (T00[i]|T10[i]|B[i]));
    R0_NXT[i]  = R0[i] | DR0[i]
               | (R0_EN & (T00[i]|T10[i]|B[i]));
    R1_NXT[i]  = R1[i] | DR1[i]
               | (R1_EN & (T01[i]|T11[i]|B[i]));
    INH_NXT[i] = INH[i] | (INH_EN & (IID == i));
    RDY_NXT[i] = RT_NXT[i] & RF_NXT[i] & R0_NXT[i]
               & R1_NXT[i] & ~INH_NXT[i];
  end
end endgenerate
\end{lstlisting}

Besides shaving LUTs, a two bank scheduler provides two sets of {\it T0, T1} ports and can sink
two sets of two events each cycle.
This is essential to sustain wider issue rates of two instructions per cycle
(which may target four operands per cycle).
Yet-wider issue and yet-larger instruction windows may even merit a four bank design.

The ready-issue-target-ready scheduling recurrence is the critical path of the IS-stage.
A 32$\rightarrow$5 priority encoder reduces the {\it RDY} vector to an {\it IID}
which selects the decoded instruction.
The decoded instruction's fields {\it \{\_T0,\_T1,\_BID,\_ENs\}} are muxed into {\it \{T0,T1,BID,ENs\}}
which update target instructions' ready state, including {\it RDY}.

{\it Priority encoder:} 
Many 32-bit encoder designs were evaluated,
including onehot conversion with LUT or carry-logic or-trees, carry-logic zero-scan, and F7MAP/F8MAP muxes.
The present design uses two 16$\rightarrow$4 encoders, one per bank, which complete in two LUT delays.
In a one-issue processor, a subsequent 2:1 mux selects one of these encoder outputs.

In particular each 16-bit encoder input {\it I[15:0]} is chunked into {\it I[15], I[14:10], I[9:5], I[4:0]}.
Each 5-bit group indexes a 32x4 LUT ROM with the precomputed encoder output for that group.
Together with three 5-bit zero comparator outputs, these feed a custom 4-bit 3:1 selector which outputs 'b1111 when
all three groups are zero.

{\it Technology mapping and floorplanning:}
The design uses an RPM (relationally placed macro) methodology to improve area and interconnect delays
and achieve a repeatable layout for easy routing and timing closure under module composition
and massive replication.
Structural RTL instantiates modules and tiles them into a scheduler.
The XST annotation {\it (*LUT\_MAP="yes"*)} on a $\le$6-input module locks its logic to one LUT;
{\it (*RLOC="XxYy"*)} packs FPGA primitives into clusters and places clusters relative
to each other.
\begin{figure}[ht]
\centerline{
  \includegraphics[width=3in,clip=true,trim=0.4in 0in 0in 0in, height=3in]{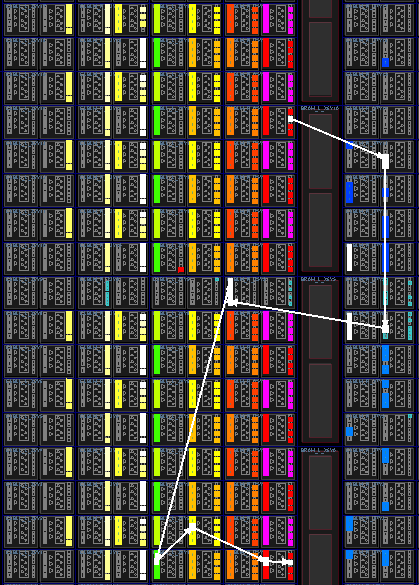}
}
\caption{FPGA implementation of the parallel scheduler}
\label{fig:par-fplan}
\end{figure}

Figure~\ref{fig:par-fplan} is a Xilinx 7-series implementation of Figure~\ref{fig:parallel},
including the scheduler, priority encoder, and decoded instruction buffer,
with the critical path highlighted in white.

Each two horizontal rows of FPGA slices correspond to four entries in the instruction window.
Left to right are:
\begin{itemize}
\item {\it pale yellow:} four 6-bit decoded ready state flip-flops;
\item {\it yellow/green:} {\it B, T00, T01, T10, T11} target decoders;
\item {\it orange:} active ready state LUTs/FFs {\it RT\_NXT/RT, etc.};
\item {\it purple:} {\it INH\_NXT} and {\it INH};
\item {\it red:} {\it RDY\_NXT} and {\it RDY}.
\end{itemize}
To the right are the synthesized priority encoders and muxes (blue) and the decoded instructions
buffer (white) implemented in several 32x6-bit true dual port LUT RAMs.

{\it Performance:} In a Kintex-7 -1-speed grade, the critical path takes 5.0 ns, including
{\it RDY} clock-to-out, priority encoder, mux, decoded instructions LUT RAM, next readys logic
and {\it RDY} setup.
Interconnect delay is 85\% of the critical path -- unfortunately all paths from any {\it RDY} to any {\it RDY}
must traverse a relatively large diameter netlist.

Cycle time may be reduced to 2.9 ns by adding a pipeline register halfway through the scheduler critical path
(the output port of the instruction buffer LUT RAM) however this will not achieve back-to-back issue
(in successive cycles) of a single dependent chain of instructions.

\subsection{Incremental dataflow scheduler ready state}
The parallel scheduler is straightforward
but it consumes hundreds of LUTs and FFs just to maintain 32x12b of ready state
-- a few LUTs worth of LUT RAM -- and this area doubles as the instruction window size doubles.
Also, each cycle its {\it next readys} LUTs recompute the readiness of every instruction,
even though (broadcast notwithstanding) each issued instruction affects
{\it at most two} others' ready state.
In contrast, the incremental scheduler keeps decoded and active ready state in LUT RAM,
maintains the frontier of ready instructions in {\it queues,}
and evaluates the ready status of just 2-4 target instructions per cycle.
Compared to an array of FFs, LUT RAM is fast and dense
but has some shortcomings:
there is no way to flash clear it, and it only supports one write per cycle.

\begin{figure}[ht]
\centering
\begin{subfigure}[b]{0.5\textwidth}
  \centering
  \includegraphics[width=2.8in, trim=0.4in 0.2in 0.5in 0.3in]{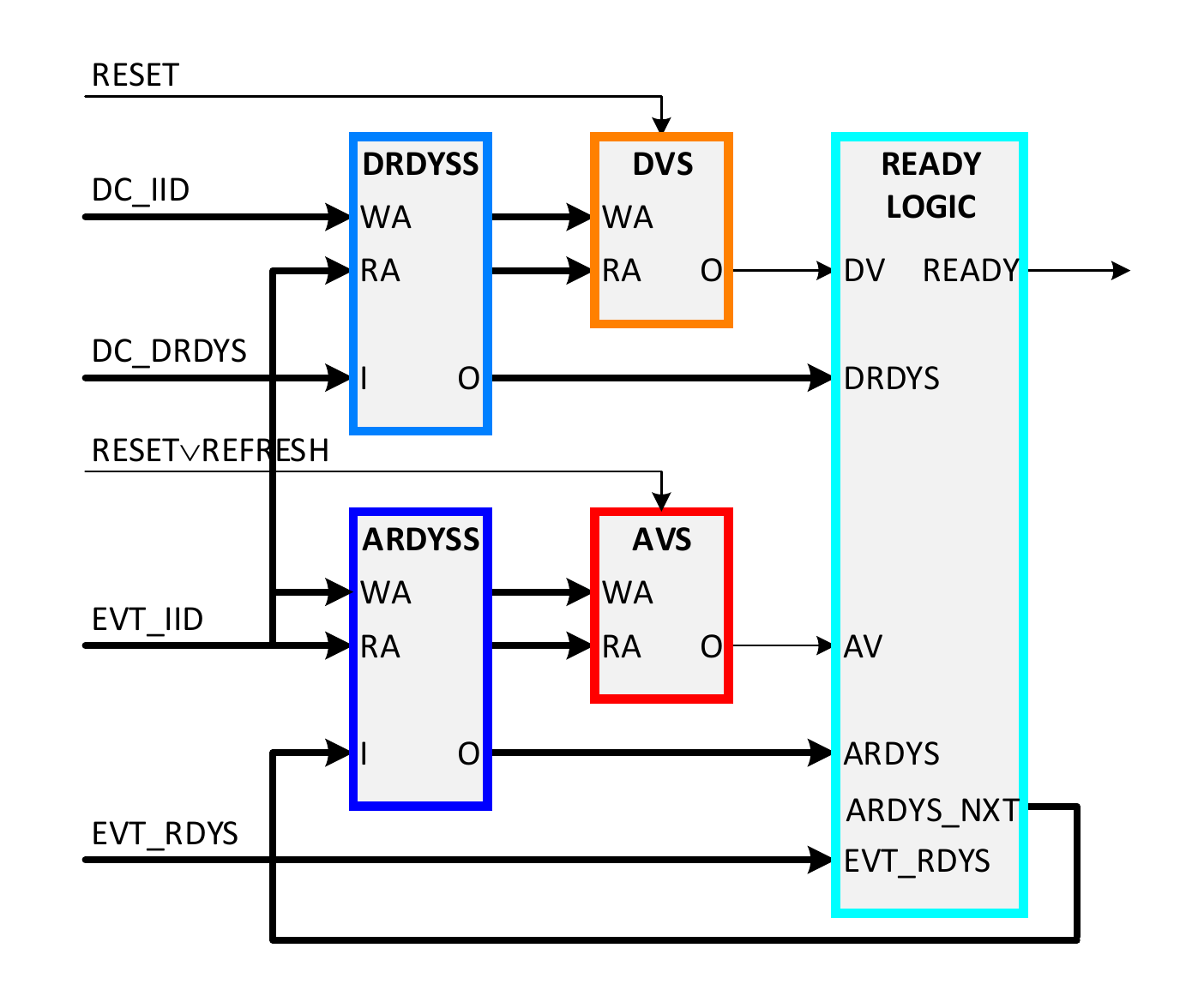}
  \caption{Design: ready state, validation, and ready logic.}
  \label{fig:incr-bank-design}
\end{subfigure}
\begin{subfigure}[b]{0.5\textwidth}
  \centering
  \includegraphics[width=2.8in]{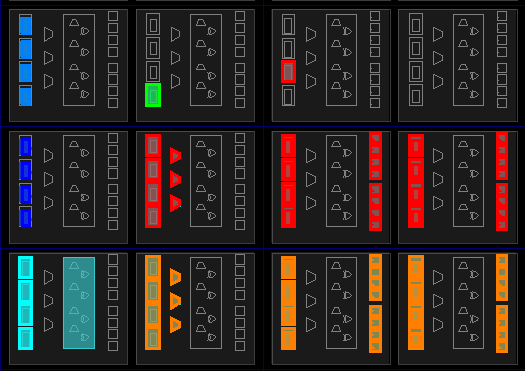}
  \caption{FPGA implementation.}
  \label{fig:incr-bank-fpga}
\end{subfigure}
\caption{A 16-entry scheduler bank.}
\label{fig:incr-bank}
\end{figure}

Instead, the scheduler uses a hybrid of LUT RAM and FF ``RAM''.
Decoded {\it (DRT, DRF, DR0, DR1)} and active {\it (RT, RF, R0, R1)} ready state are
stored in several banks of 16x4 true dual port LUT RAM,
which is {\it validated} by a 16x1 flash-clearable-set-only-RAM ``FC-SO-RAM''.
This comprises 16 FFs (with common reset), 16 write port address decoders (eight 5,5-LUTs),
and a 16:1 read port mux (four 6-LUTs, two MUXF7s, one MUXF8) -- just three slices in all.
Each read from this hybrid reads the 4b LUT RAM entry and its valid bit.
Each write updates the LUT RAM and sets its valid bit.

{\it Multiple LUT RAM write ports.}
To sustain a fetch/decode rate of {\it d} instructions/cycle,
and an issue rate of {\it i} instructions/cycle, it is necessary to update
{\it d+2i} ready state entries each cycle.
This is a challenge for one write/cycle LUT RAM.
Rather than use clock doubling or replicated RAM banks with live value tables,
the incremental scheduler divides the ready state up into four (or more) interleaved,
disjoint banks:
(decoded, active) ready state for (even, odd) instructions.
Then the front end can write even and odd decoded ready state
while the back end updates even and/or odd target instructions' active ready state.

Figure~\ref{fig:incr-bank} shows the resulting a 16-entry scheduler
bank design and implementation.
The blue decoded and active ready state LUT RAMs {\it DRDYSS} and {\it ARDYSS} are validated
by orange/red FC-SO-RAMs {\it DVS} and {\it AVS}.
Each cycle the decoder writes instruction {\it DC\_IID}'s decoded ready state {\it DC\_DRDYS}
and its valid bit.
Also each cycle the bank's target ready event {\it EVT ::=\{EVT\_IID; EVT\_RDYS\}} is processed
via a read-modify-write of {\it EVT\_IID}'s {\it ARDYS} with its {\it DRDYS} and {\it EVT\_RDYS}.
See Listing~\ref{readylogic}.
The instruction is ready when all four {\it ARDYS} bits are set. 
All of this logic (cyan) requires just one slice; as an optimization
{\it READY's and-}reduction is in carry-logic.

\begin{lstlisting}[label=readylogic,caption=Ready logic]
// ready logic
always @* begin
  ARDYS_NXT = (DV ? DRDYS : 4'b0000)
            | (AV ? ARDYS : 4'b0000)
            | EVT_RDYS;
  READY = &ADRYS_NXT;
end
\end{lstlisting}

Note that the EDGE compiler does not guarantee that both targets of an
instruction are in disjoint banks so there may be scheduler bank
conflicts.  An ADD instruction might target an operand of instruction
10 and an operand of instruction 12.  Since it is not possible to
update the active ready state of the two even bank targets in the
same cycle, one event is processed and the other is queued for a later cycle.

\subsection{Incremental dataflow scheduler design, operation, and implementation}
\begin{figure}[bht]
\centering
\begin{subfigure}[b]{0.5\textwidth}
  \centering
  \includegraphics[width=3in, trim=0.4in 0.3in 0.4in 0.2in]{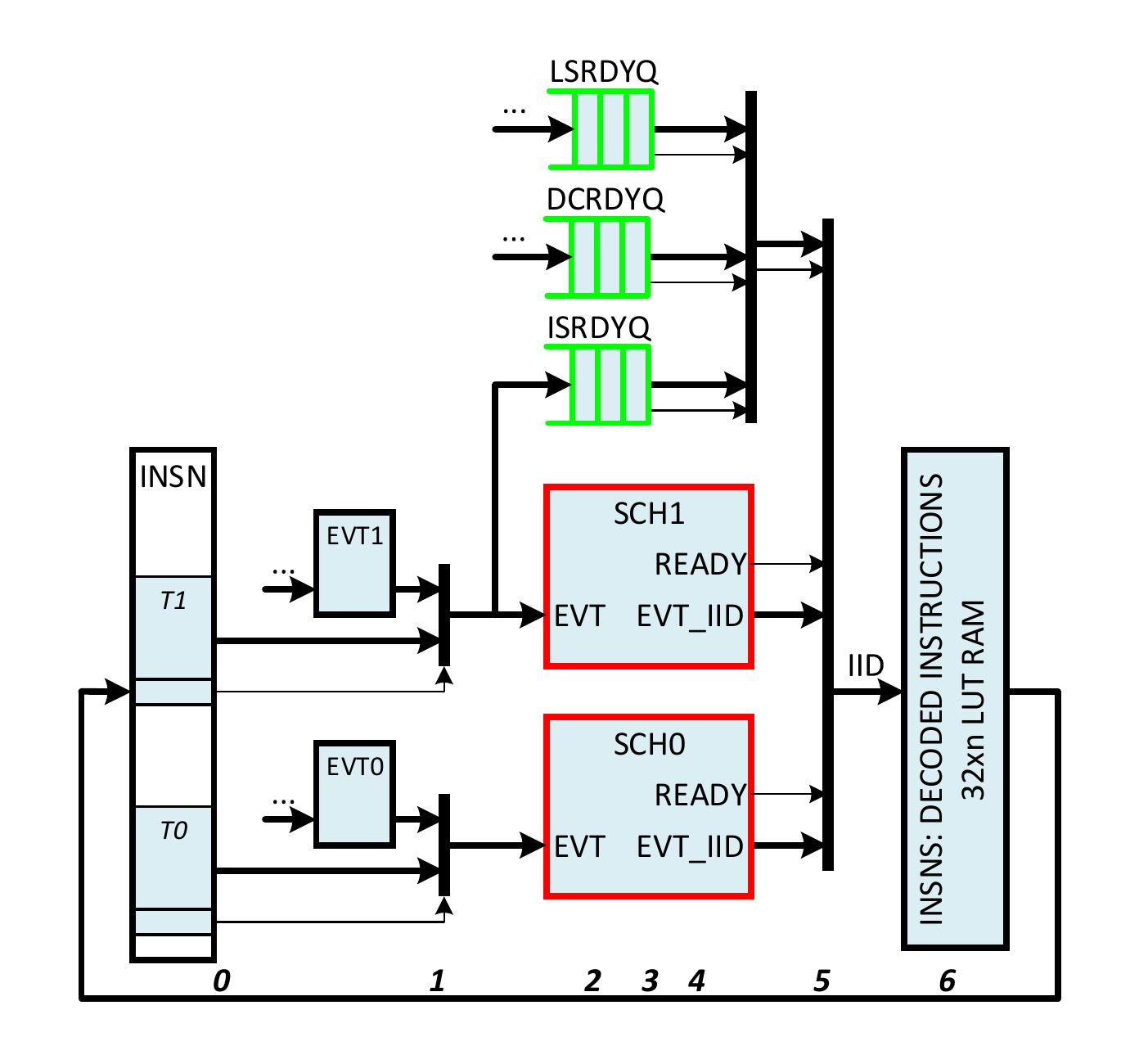}
  \caption{Design.}
  \label{fig:incr-sched-design}
\end{subfigure}
\begin{subfigure}[b]{0.5\textwidth}
  \centering
  \includegraphics[width=2.8in]{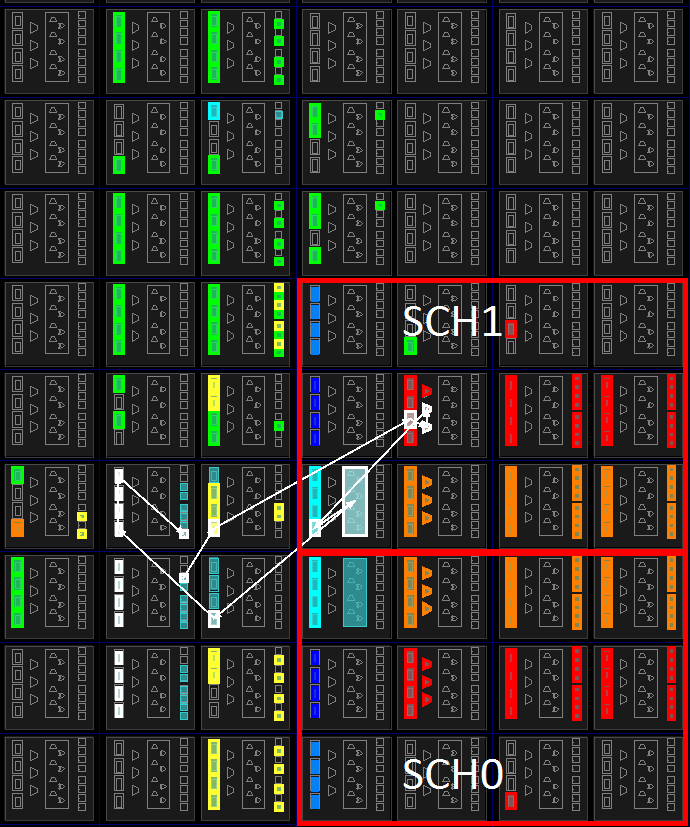}
  \caption{FPGA implementation.}
  \label{fig:incr-sched-fpga}
\end{subfigure}
\caption{32-entry scheduler, decoded instructions buffer, and ready queues.}
\label{fig:incr-sched}
\end{figure}
The core of the scheduler (Figure~\ref{fig:incr-sched}) consists of:
\begin{itemize}
  \item {\it INSN:} decoded instruction with two target event fields
  \item {\it EVT0, EVT1:} even/odd pending event registers
  \item even/odd event muxes, controlled by predecoded selects
  \item {\it SCH0, SCH1:} even/odd 16-entry scheduler banks
  \item three ready instruction {\it IID} queues:
  \begin{itemize}
    \item {\it DCRDYQ:} decoder ready queue;
    \item {\it ISRDYQ:} issue (scheduler) ready queue;
    \item {\it LSRDYQ:} load/store ready queue
  \end{itemize}
  \item two 3:1 selectors to select next {\it IID}
  \item {\it INSNS:} decoded instructions RAM (read port)
\end{itemize}
Note, in this design, the scheduler recurrence cycle begins and ends with the
decoded instruction register.

Now consider execution of the first EDGE code block in Figure~\ref{fig:edge}.
The scheduler is reset, clearing {\it DVS, AVS} in {\it SCH0, SCH1}.
The front end fetches the block's header and
fetches and decodes its instructions into {\it INSNS}.
The two READs are ready to issue so their {\it IIDs} are enqueued on {\it DCRDYQ}.
This ``primes the pump'' for the back end.
The other instructions await operands or predicates, are not ready, so are not enqueued.

Back end dataflow execution proceeds as follows.
Initially {\it INSN} is invalid and both {\it READYs} are negated.
The {\it IID} selector tree selects/dequeues the first READ
instruction {\it (IID=0)} from {\it DCRDYQ}.
The decoded READ instruction word is read from {\it INSNS} into {\it INSN}.

The READ targets ADD operand \#1.
Its {\it INSN.T0} (even bank target ready event) field is valid and
its mux selects {\it EVT}=(2,'b0001) for {\it SCH0}.
That updates ADD's active ready state: 'b1100\textbar 'b0000\textbar 'b0001='b1101,
now awaiting only the left operand (operand \#0).
Since neither scheduler bank found a {\it READY} instruction,
the {\it IID} selector tree selects/dequeues the second READ from {\it DCRDYQ}.

This READ targets ADD operand \#0; its {\it INSN.T0} is {\it EVT}=(2,'b0010).
{\it SCH0} updates ADD's ready state to {'b1111} and asserts {\it READY}
so the ADD {\it (IID=2)} issues.

ADD's {\it T1} targets the TLEI ready state in {\it SCH1}.
TLEI becomes ready and issues.

As for TLEI, neither {\it T0/T1} fields designate IS-stage ready events.
Why?
Unlike simple one cycle latency instructions like ADD, test instructions' targets
cannot receive ready events until the test executes in the EX-stage.
Once a test completes, its true/false predicate event(s) are signaled.
These proceed through queues and/or muxes (not shown) to the {\it EVT0, EVT1}
pending event registers, awaiting idle scheduler event slots.


{\it Queues:}
The design employs many elastic FIFO ready queues and event queues.
They are small and fast, built with up-down counters and
Xilinx SRL32CE 32-bit variable length shift register LUTs.
Besides {\it DCRDYQ} the present design has two other ready queues.

{\it ISRDYQ:}
In a ``one issue'' design
when an instruction issues and it targets and wakes two others,
the even instruction issues next, and the odd one is queued on {\it ISRDYQ}.

{\it LSRDYQ:}
EDGE processors use a load-store queue to provide sequential memory semantics.
One simple area-optimized LSQ defers and reorders certain accesses;
then when a (ready) load/store becomes issuable {\it to memory} the LSQ enqueues it on {\it LSRDYQ}.

{\it Broadcast wakeup:}
Each EDGE result broadcast may target and wake an arbitrary number of instructions
in the window.
This is easy for a parallel scheduler but costly for an incremental one.
When a result is broadcast the scheduler must sequentially update
the ready state of every decoded instruction with that broadcast input.
Therefore the decoder maintains queues {\it (BR1Q, BR2Q, BR3Q)}
of {\it IIDs} of instructions with a given broadcast input.
Once a broadcast result is known, the scheduler begins to dequeue the {\it BRnQ IIDs}
into {\it EVTs} presented to {\it SCH0, SCH1}.

{\it Performance:}
The labels {\it 0-6} in Figure~\ref{fig:incr-sched-design}
depict the number of ``LUT delays'' to each point in the scheduler critical path,
the white path in Figure~\ref{fig:incr-sched-fpga}.
In a Kintex-7 -1-speed grade, this takes 4.3 ns
including {\it INSN} clock-to-out, {\it EVT} mux, {\it SCH1's AVS} read port mux,
{\it ARDYS\_NXT} and {\it READY} logic, {\it IID} selector,
{\it INSNS} read, and {\it INSN} setup.
Here interconnect delay is just 70\% of the critical path reflecting relatively
shorter nets and some use of LUT-local MUXF7/MUXF8/CARRY4 nets.
The scheduler clock period may be reduced to 2.5 ns by adding pipeline registers
after the LUT RAM and FC-SO-RAM reads, but as with the parallel scheduler,
pipelining precludes back-to-back issue of dependent instructions.

\subsection{Comparing parallel and incremental schedulers}
\begin{table}[ht]
\centering
\begin{tabular}{|l||c|c|l|}
\hline
Metric&Parallel&Incremental&Units\\
\hline
Area, 32 entries&288&78&LUTs\\
Area, total, 32 entries&340&150&LUTs\\
Period&5.0&4.3&ns\\
Period, pipelined&2.9&2.5&ns\\
Area, total * period&{\bf 1700}&{\bf 645}&LUT*ns\\
\hline
Broadcast&flash&iterative&\\
Event bank conflicts?&never&sometimes&\\
\hline
Area, 4 events/cycle&288&156&LUTs\\
Area, 64 entries&576&130&LUTs\\
\hline
\end{tabular}
\caption{Comparing parallel and incremental schedulers.}
\label{tab:par-incr}
\end{table}

Table~\ref{tab:par-incr} summarizes the differences between the two dataflow scheduler designs.
The core of the incremental scheduler is less than a third the size of the parallel scheduler
although the size advantage is smaller when the additional overhead of queues and muxes is added.
The incremental scheduler is also faster and the  area*period metric is 2.6x better.
However the parallel scheduler retains some brute force advantages.
It can process a broadcast event in a single cycle,
whereas the incremental scheduler must iteratively drain a broadcast queue
at a rate of 1-2 instructions per cycle.
This may cause issue stalls in some workloads.
The incremental scheduler is also subject to even/odd target bank conflicts
which may delay an instruction wake up.
Real workload studies are needed to measure whether these effects overshadow
its substantial area*period advantage.

Finally consider future scale up to wider issue and larger instruction windows.
The parallel scheduler does not grow when subdivided into more banks
to process twice as many events per cycle,
whereas the incremental scheduler core area doubles.
To grow the instruction window to 64 entries,
the parallel scheduler require twice as much area,
whereas the incremental scheduler area grows more modestly.

\section{Conclusion}
This paper presents our work towards a practical
out-of-order dataflow soft processor architecture for FPGAs.  We
set out to discover whether a novel EDGE instruction set architecture,
optimized for simpler high ILP microarchitectures in ASICs, is also
a good fit for FPGAs, or whether general purpose soft processors
will remain stuck in the scalar RISC slow lane.

We considered two different dataflow instruction scheduler designs
and their optimized FPGA implementations.  In the context of
commercial 200 MHz, 1,000-2,000 LUT soft processors, the limited
FPGA resource cost and clock period impact of either design seems
acceptable and practical.  Both design alternatives will scale well
to future four-decode/two-issue implementations.


\bibliographystyle{IEEEtran}
\bibliography{paper}

\end{document}